\documentclass[number,preprint,review,3p,12pt]{elsarticle}

\usepackage{amsmath,amsfonts,mathrsfs,tabu}
\usepackage{booktabs}
\usepackage{color}
\usepackage[ngerman,english]{babel}
\usepackage{graphicx}
\usepackage{epstopdf}
\usepackage{caption}
\usepackage{subcaption}
\usepackage{siunitx}
\usepackage{multirow}
\usepackage[final]{pdfpages}

\usepackage{url}
\usepackage[labelformat=simple]{subcaption}

\usepackage{dblfloatfix}
\usepackage{graphicx}
\usepackage{epstopdf}
\usepackage{xcolor}
\usepackage{mathtools,amssymb,lipsum}
\usepackage[inline]{enumitem}
\usepackage{cuted}
\usepackage[figure]{algorithm2e} 
\usepackage{float}
\usepackage[inline]{enumitem}

\usepackage{makecell}

\newcommand{\sas}{S$\alpha$S}
\newcommand{\highlight}[1]{\textcolor{blue}{#1}}
\renewcommand{\highlight}[1]{#1}

\journal{Physical Communication}

\begin{document}

\begin{frontmatter}



\title{\highlight{Merging} the Bernoulli-Gaussian and Symmetric $\alpha$-Stable Models for Impulsive Noises in Narrowband Power Line Channels}


 \author[tukladdress]{Bin~Han\corref{mycorrespondingauthor}}
 \cortext[mycorrespondingauthor]{Corresponding author}
 \ead{binhan@eit.uni-kl.de}
 \author[geiriaddress]{Yang~Lu}
 \ead{luyang@geiri.sgcc.com.cn}
 \author[geiriaddress]{Kai~Wan}
 \ead{wankai@geiri.sgcc.com.cn}
 \author[tukladdress,dfkiaddress]{Hans~D.~Schotten}
 \ead{schotten@eit.uni-kl.de}


\address[tukladdress]{Institute of Wireless Communication (WiCon), Technische Universit\"at Kaiserslautern, 67663 Kaiserslautern, Germany}
\address[dfkiaddress]{Research Group Intelligent Networks, German Research Center for Artificial Intelligence (DFKI GmbH), 67663 Kaiserslautern, Germany}
\address[geiriaddress]{Global Energy Interconnection Research Institute (GEIRI), State Grid Corporation of China (SGCC), Beijing 102209, China}

\begin{abstract}
To model impulsive noise in power line channels, both the Bernoulli-Gaussian model and the symmetric $\alpha$-stable model are usually applied. Towards a merge of  existing noise measurement databases and a simplification of communication system design, the compatibility between the two models is of interest. In this paper, we show that they can be approximately converted to each other under certain constrains, although never generally unified. Based on this, we propose a fast model conversion.
\end{abstract}


\end{frontmatter}


\section{Introduction}\label{sec:intro}
Impulsive noise, which is generated by numerous electrical devices connected to the power grid, ubiquitously exists in power line channels. These noise peaks with high amplitude can erase most communication signals transmitted over the power line channel, and are proved  to significantly impact the performance of power line communication (PLC) systems. Meanwhile, the occurrence of such impulses is challenging to predict because of its highly non-stationary dynamics. An intensive interest in modeling noises of this type therefore arises, driven by the demand of impulsive noise mitigation for PLC. Since over two decades, different time-domain models have been proposed or adopted to characterize them, including:
	\begin{itemize}
		\item the Middleton's Class-A (MCA) model~\cite{middleton1979procedures}, which characterizes the sparsity of high-amplitude spikes in noise;
		\item the Bernoulli-Gaussian (BG) model~\cite{ghosh1996analysis}, which considers the impulsive noise as a Bernoulli sequence modulated to a Gaussian noise with On-Off-Keying;
		\item the Symmetric $\alpha$-Stable (\sas) model~\cite{laguna2014experimental}, which statistically describes the distribution of noise amplitude;
		\item the Markov-Middleton model~\cite{ndo2013markov} , which is an extended MCA model where the MCA parameters randomly switch among several states;
		\item the Markov-Gaussian model~\cite{fertonani2009reliable}, which extends the BG model by replacing the Bernoulli process with a Markov process.
\end{itemize}

Comparative studies on these models have been reported~\cite{shongwe2014impulse,han2017noise}. Generally, the Markov-Middleton and Markov-Gaussian models are enhanced variations of the MCA and BG models, respectively, which introduce Markov chains to describe the burst noise phenomenon. When ignoring noise bursts, the MCA model, the BG model and the \sas~model are mainly used. As pointed out in \cite{shongwe2014impulse}, the BG model is usually preferred over the MCA model for its better tractability. Meanwhile, focusing on the statistics instead of the dynamics of noise, the \sas~model outperforms the MCA model with its excellent performance in fitting the heavy-tailed probability density function (PDF) of noise amplitude. 

Comparing the BG model with the \sas~model, each side has its own pros and cons. On the one hand, the BG model is \highlight{cost-friendly for implementations}, and can be easily \highlight{extended} to the Markov-Gaussian model to describe burst noise. In contrast, the \sas~model cannot model burst noise, and is \highlight{expensive to compute due to the lack of generic close form PDF}. On the other hand, the \sas~model can accurately match the fat-tailed distribution of impulsive noises, and the parameters can be consistently estimated from the amplitude statistics~\cite{tsihrintzis1996fast}. The BG model, in comparison, does not guarantee a good fit for the overall sample amplitude distribution, and its parameter estimation highly relies on the accuracy of impulse detection and extraction~\cite{han2017noise}.

In PLC system design, upon specific requirements of different applications, one or several models listed above can be preferred over the others and therefore flexibly selected. \highlight{For example, when it is essential to consider impulsive noises with broader bandwidth than the signals, such like in cognitive PLC systems that flexibly select the working frequency range~\cite{liu2014power}, the \sas~model can be preciser~\cite{shongwe2014impulse}. In contrast, for cost-critical narrowband PLC applications such as smart grid systems~\cite{galli2011grid}, the BG model can be more practical as a high computational effort is required for the \sas~model.} However, from the perspectives of channel measurement and system evaluation, there is a solid demand of unification or conversion between different noise models. First, field measurements of power line noises are usually expensive in cost and effort, and the measured results are usually reported and archived in the form of estimated model parameters instead of raw data, e.g. as it is done in \cite{cortes2010analysis}. A unification among various noise models will enable to reuse the valuable data of measurements in different applications and thereby greatly save the measuring cost. Moreover, in the evaluation of PLC systems, to ensure the generality of results, it is often necessary to repeat the test with \highlight{noises generated by different models}, as reported in \cite{lin2011non}. A noise model unification also helps reduce such effort by a significant degree.
	
Fortunately, the Markov-Middleton and Markov-Gaussian models are endogenously compatible with the MCA and BG models, respectively. Meanwhile, the BG model has been demonstrated as capable to approximate the MCA model with simple adjustments~\cite{shongwe2014impulse}. However, the compatibility between \highlight{the} BG and \sas~models, to the best of our knowledge, has never been throughly investigated yet. Focusing on this unsolved problem, in this paper we: \begin{enumerate*}
		\item demonstrate the similar performance of  these two models in characterizing impulsive power line noises,
		\item derive the quasi-stability of BG noise in \highlight{the} context of PLC, and
		\item propose a fast and approximate polynomial conversion from the \highlight{the} model to \highlight{the} \sas~model.
	\end{enumerate*}

The remainder of this manuscript is organized as follows. We review the BG and \sas~models in Section \ref{sec:model_review}. Then we comparatively evaluate both \highlight{of them} with \highlight{filed measurements of power line noise} in Section \ref{sec:experiment}. Subsequently, in Section \ref{sec:stability_bg} we analyze the compatibility between the \highlight{two models}. Our results indicate that they cannot not be generally unified, but are compatible with each other to a \highlight{satisfactory} degree, especially when the impulses are sparse and limited in power. Afterwards, in Section \ref{sec:sas_para_of_bg}, we invoke existing signal processing techniques to fit BG processes with the \sas~model, in order to build a polynomial \highlight{model of conversion} between BG and \sas~parameters, and evaluate its fitting performance. The details of all experimental results are aggregated in Section \ref{sec:results}. At the end we close this paper with our conclusions in Section \ref{sec:conclusion}.

\section{Bernoulli-Gaussian Model and S$\alpha$S Model}\label{sec:model_review}
\subsection{Bernoulli-Gaussian Model}
The BG model describes a sampled impulsive noise as
\begin{equation}
n_\textrm{I}(k)=\sigma_\textrm{I}^2n_\textrm{G}(k)\phi(k)
\end{equation}
where $k\in\mathbb{Z}$ is the sample index, $\sigma_\textrm{I}$ is the standard deviation of the impulsive noise amplitude and $n_\textrm{G}(k)$ is a normalized white Gaussian noise with \highlight{a} unity power. $\phi(k)$ is a Bernoulli process that describes the occurrence of impulses:
\begin{equation}
\phi(k)=\begin{cases}
1&z(k)\le p\\
0&\textrm{otherwise}
\end{cases},
\end{equation}
where $z(k)\sim U(0,1)$ and $p\in[0,1]$ is the impulse probability.
In practice, it is usual to consider the mixture of impulsive noise and Gaussian background noise as
\begin{equation}
n_\textrm{BG}(k)=\sigma_\textrm{B}^2n_0(k)+\sigma_\textrm{I}^2n_1(k)\phi(k),
\label{equ:bg_model}
\end{equation}
where $\sigma_\textrm{B}$ is the standard deviation of the background noise amplitude, $n_0(k)$ and $n_1(k)$ are two {independent} Gaussian noises with unity power. Thus, the PDF of $n_\textrm{BG}$ is
\begin{equation}
f_{n_\textrm{BG}}(x)=\frac{1-p}{\sqrt{2\pi\sigma_\textrm{B}^2}}\mathrm{e}^{-\frac{x^2}{2\sigma_\textrm{B}^2}}+\frac{p}{\sqrt{2\pi(\sigma_\textrm{B}^2+\sigma_\textrm{I}^2)}}\mathrm{e}^{-\frac{x^2}{2(\sigma_\textrm{B}^2+\sigma_\textrm{I}^2)}}.
\label{equ:bg_pdf}
\end{equation}

\subsection{Symmetric $\alpha$-Stable Model}
A random variable $X$ is called \textit{stable} if and only if 
\begin{equation}
\exists a\in\mathcal{R}^+, b\in\mathcal{R}^+,c\in\mathcal{R}^+, d\in\mathcal{R}: aX_1+bX_2\stackrel{\mathcal{D}}{=}cX+d,
\label{equ:stable_process_def_1}
\end{equation}
where $X_1$ and $X_2$ are two \textit{independent} copies of $X$ and $A\stackrel{\mathcal{D}}{=}B$ denotes that $A$ and $B$ obey the same \highlight{statistical} distribution. Especially, the distribution is called \textit{strictly stable} if \highlight{\eqref{equ:stable_process_def_1}} holds \highlight{for} $d=0$~\cite{nolan2017stable}.

\highlight{The definition above is proved to have the following equivalence}: $X$ is $\alpha$-stable if and only if $\exists (0<\alpha\le2,-1\le\beta\le1),\gamma\neq0,\delta\in\mathcal{R}$ that
\begin{equation}
X\stackrel{\mathcal{D}}{=}\gamma Z+\delta,
\end{equation}
where $Z$ is a random variable with characteristic function
\begin{equation}
\varphi(t)=\begin{cases}
\mathrm{e}^{{j\delta t-\gamma|t|^\alpha\left[1-i\beta\tan\frac{\pi\alpha}{2}\mathrm{sign}(t)\right]}}&\alpha\neq 1\\
\mathrm{e}^{{j\delta t-\gamma|t|\left[1+i\beta\frac{2}{\pi}\mathrm{sign}(t)\log|t|\right]}}&\alpha=1
\end{cases}.
\end{equation}
\highlight{Usually we refer to $\alpha$ as the {index of stability} or {characteristic exponent}, $\beta$ as the skewness parameter, $\gamma$ as the scale parameter and $\delta$ as the location parameter}~\cite{nikias1995signal}. Especially, when $\beta=0$, the PDF of $X$ is symmetric about $\gamma$\highlight{, and} $X$ is called {symmetric $\alpha$-Stable} (\sas). Some special cases of $\alpha$-stable distribution have simple expressions of PDF, and have been well-studied, including $\alpha=2$, $\beta=0$ (Gaussian); $\alpha=1$, $\beta=0$ (Cauchy) and $\alpha=0.5$, $\beta=1$ (L\'evy). However, field measurements have proved that when applied on PLC noises, the $\alpha$-stable model usually has parameters $\alpha\in(1.5,2)$, $\beta\approx 0$ ~\cite{laguna2015use}, which is a \sas~case without any closed-form presentation of PDF.

\section{Test with Field Measurements}\label{sec:experiment}
\begin{figure}[!htbp]
	\centering
	\includegraphics[width=.6\textwidth]{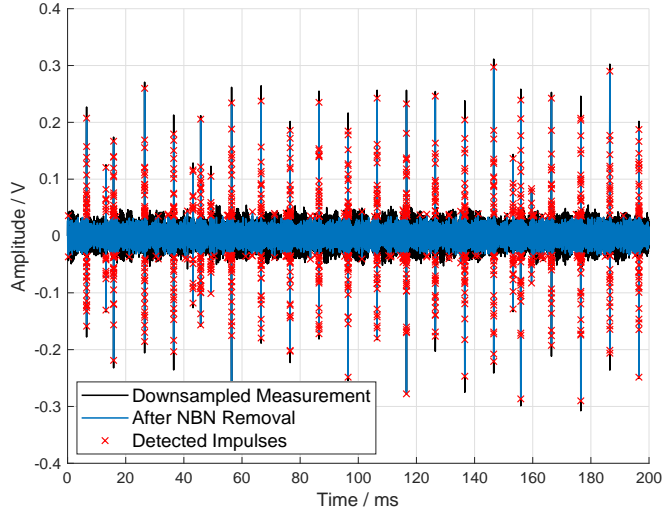}
	\caption{A sample fragment of the noise measurement with preprocessing results.}
	\label{fig:sample_noise}
\end{figure}

To evaluate the performance of both models, we \highlight{test} them with filed measurements. The raw data \highlight{were captured} in 2012 at the B-phase live wire on the secondary side of a low-voltage transformer\highlight{, which was} located in the power distribution room of a urban residential district in China. The measurement was executed twice, once on September 9\textsuperscript{th} at 18:17 and the other on September 14\textsuperscript{th} at 01:08, each \highlight{lasting} \SI{200}{\milli\second} with the sampling rate of 80 MSPS.
Instead of working with the raw measurement, we \highlight{down-sample} the data to 2 MSPS with a Butterworth anti-aliasing filter, due to two reasons: 
\begin{enumerate}
	\item The simple BG model is supposed to \highlight{be applied} on impulsive noises in underspread channels, where the impulse width is significantly shorter than the sampling interval, so that the multi-path channel fading can be ignored~\cite{han2018fast}. Under a very high sampling rate such as 80 MSPS, such approximation \highlight{does not hold any more,} and the noise must be first de-convoluted from an unknown observation matrix before \highlight{fitted with} the BG model, which would complicate the task.
	\item Through the downsampling, the data size and hence the computational cost are reduced. Meanwhile, both the BG and \sas~models are consistent to downsampling, so that their performance will not be impacted.
\end{enumerate}
As indicated in~\cite{han2017noise}, high-powered narrowband interferers in PLC \highlight{systems} are usually amplitude modulated by periodical envelopes \highlight{synchronous} to the mains voltage, and can \highlight{thus} exhibit deterministic impulsive behavior. Therefore, they may significantly interfere the analysis of stochastic impulsive components in noise. Here we \highlight{invoke} the Narrowband Regression method~\cite{han2015novel} to cancel periodically fluctuating narrowband interferers from the down-sampled \highlight{measurements}. Then we \highlight{invoke} the blind BG impulse detector reported in~\cite{han2018fast} on the cleaned \highlight{results} to distinguish \highlight{the spikes} from the background noise. An instance of noise preprocessing \highlight{result} is depicted in Fig.~\ref{fig:sample_noise}. For the BG model, the impulse ratio $p$ and the Gaussian parameters $\left(\sigma_1^2,\sigma_2^2\right)$ can be easily estimated from the labeled data. For the \sas~model, we applied {McCulloch}'s method~\cite{mcculloch1986simple} for parameter estimation. 

Then, based on the estimated model parameters, we numerically \highlight{generate} BG and \sas~noise sequences, each with a length of $5\times10^5$ samples. Subsequently, we \highlight{compare} the empirical PDFs of measured and simulated noise \highlight{amplitudes}. To evaluate the fitnesses of both models, we further \highlight{calculate} the weighted root mean square error (RMSE) for each of them:
\begin{equation}
	\epsilon_\text{model}=\sqrt{\int_{-\infty}^{+\infty}f_\text{meas}(x)[f_\text{model}(x)-f_\text{meas}(x)]^2\text{d}x},
\end{equation}
where $f_\text{meas}$ and $f_\text{model}$ are the empirical PDFs of \highlight{measured and simulated noise amplitudes}, respectively. The results, which are \highlight{detailed} in Section~\ref{subsec:results_plc_test}, demonstrate that both models exhibit fitting performance to a similar degree \highlight{of satisfaction}. 

\section{Stability of Bernoulli-Gaussian Processes}\label{sec:stability_bg}
Towards a unification between \highlight{the} BG and \sas~models, the first question is: \textit{are BG processes stable?} To answer this, we \highlight{test} if the BG model defined in \eqref{equ:bg_model} fulfills the requirement of stability defined in \eqref{equ:stable_process_def_1}. Consider two idenpendent and identically distributed (i.i.d.) variables $X$ and $Y$ which are generated according to \eqref{equ:bg_model}, and their sum $W$:
\begin{equation}
w(k)=x(k)+y(k),\forall k\in\mathbb{Z}.
\end{equation}
The PDF of $W$ will then be
\begin{equation}
f_W(w)=\int\limits_{-\infty}^{+\infty}f_Y(w-x)f_X(x)\mathrm{d}x.
\end{equation}
As both $X$ and $Y$ have the same PDF as given in \eqref{equ:bg_pdf}, we have\footnote{For the detailed derivation see Appendix.}
\begin{align}
\begin{split}
&f_W(w)=\int\limits_{-\infty}^{+\infty}f_\textrm{BG}(w-x)f_\textrm{BG}(x)\mathrm{d}x\\
=&\frac{1-2p+p^2}{\sqrt{4\pi\sigma_\textrm{B}^2}}\mathrm{e}^{-\frac{w^2}{4\sigma_\textrm{B}^2}}+\frac{p^2}{\sqrt{4\pi(\sigma_\textrm{B}^2+\sigma_\textrm{I}^2)}}\mathrm{e}^{-\frac{w^2}{4(\sigma_\textrm{B}^2+\sigma_\textrm{I}^2)}}\\
+&\frac{p-p^2}{\sqrt{2\pi(2\sigma_\textrm{B}^2+\sigma_\textrm{I}^2)}}\left(\mathrm{e}^{-\frac{(2\sigma_\textrm{B}^2+\sigma_\textrm{I}^2-1)w^2}{2(\sigma_\textrm{B}^2+\sigma_\textrm{I}^2)(2\sigma_\textrm{B}^2+\sigma_\textrm{I}^2)}}+\mathrm{e}^{-\frac{(2\sigma_\textrm{B}^2+\sigma_\textrm{I}^2-1)w^2}{2\sigma_\textrm{B}^2(2\sigma_\textrm{B}^2+\sigma_\textrm{I}^2)}}\right).
\end{split}\label{equ:bg_sum_pdf}
\end{align}

Clearly, \eqref{equ:bg_sum_pdf} differs in form from \eqref{equ:bg_pdf}, so that we know Bernoulli-Gaussian processes are \textit{not generally stable}. Only in the following three special cases, $W$ is quasi-stable as it approximately approaches to a Gaussian process:
\begin{align}
\lim\limits_{p\to 0}f_W\left(\frac{w}{\sqrt{2}}\right)&=\lim\limits_{p\to 0} f_\textrm{BG}(w)\approx  f_\textrm{B}(w);\label{equ:quasi_stable_1}\\
\lim\limits_{p\to 1}f_W\left(\frac{w}{\sqrt{2}}\right)&=\lim\limits_{p\to 1} f_\textrm{BG}(w)\approx  f_\textrm{I}(w);\label{equ:quasi_stable_2}\\
\lim\limits_{\sigma_\textrm{I}^2-\sigma_\textrm{B}^2\to 0}f_W\left(\frac{w}{\sqrt{2}}\right)&= f_\textrm{BG}(w)\approx f_\textrm{B}(w)\approx  f_\textrm{I}(w),\label{equ:quasi_stable_3}
\end{align}
where $f_\textrm{B}(w)$ and $f_\textrm{I}(w)$ are the PDFs of $n_\textrm{B}(k)$ and $n_\textrm{I}(k)$, respectively. \highlight{The approximation \eqref{equ:quasi_stable_1} is} valid in the context of PLC, where $p$ is \highlight{sufficiently low although} $\sigma_\textrm{I}^2$ is significantly higher than $\sigma_\textrm{B}^2$, as it will be demonstrated and discussed with details in Section~\ref{subsec:results_bg_stability}.

\section{Estimating \sas~Parameters of BG Processes}\label{sec:sas_para_of_bg}
So far, we have \highlight{demonstrated} a certain but limited  compatibility between the BG and \sas~models for power line noises. Subsequently, driven by the interest in the performance of approximating BG models with \sas~models, we \highlight{attempt} to apply the \sas~model on BG processes.

Our methodology can be summarized as follows. First, we \highlight{generate} a BG noise with parameters $(p,\sigma_\textrm{B}=1,\sigma_\textrm{I})$, and \highlight{normalize} it to  $n_\textrm{BG}$ with unity power. Then we \highlight{apply} {McColloch}'s \sas~parameter estimator~\cite{mcculloch1986simple} on it. Subsequently, we \highlight{call} {Chambers}' method~\cite{chambers1976method} to simulate a \sas~noise sequence $n_\textrm{\sas}$ with the estimated parameters $(\hat{\alpha},\hat{\gamma})$. Afterwards, we \highlight{evaluate} the fitting performance with the Kullback-Leibler divergence ~\cite{kullback1997information}:
\begin{equation}
D_\textrm{KL}(\phi_\textrm{BG}|\phi_\textrm{\sas})=\int\limits_{-\infty}^{+\infty}\phi_\textrm{BG}(x)\ln\frac{\phi_\textrm{BG}(x)}{\phi_\textrm{\sas}(x)}\mathrm{d}x,
\end{equation}
where $\phi_\textrm{BG}$ and $\phi_\textrm{\sas}$ denote the empirical PDF of $n_\textrm{BG}$ and $n_\textrm{\sas}$, respectively. This divergence, also known as the relative entropy, is normalized to the range $[0,1]$ and measures the degree that $\phi_\textrm{BG}$ diverges from $\phi_\textrm{\sas}$. $D_\textrm{KL}(\phi_\textrm{BG}|\phi_\textrm{\sas})=0$ indicates that both the distributions are highly similar, if not the same, while $D_\textrm{KL}(\phi_\textrm{BG}|\phi_\textrm{\sas})=1$ denotes a minimal similarity between the distributions.
By repeating this process with different Bernoulli-Gaussian parameters, we are able to investigate the dependencies of $(\hat{\alpha},\hat{\gamma})$ on $(p,{\sigma_\textrm{I}^2}/{\sigma_\textrm{B}^2})$.

Due to the lack of closed-form density functions, most conventional analytic methods of statistics cannot be applied to estimate \sas~parameters for cases where $\alpha>1$. Nevertheless, a variety of numerical techniques have been developed for this task. According to ~\cite{nikias1995signal}, classical approaches can be generally classified into four categories: methods of maximum likelihood~\cite{brorsen1990maximum,nolan2001maximum}, methods of sample fractiles~\cite{mcculloch1986simple,tsihrintzis1996fast}, methods of sample characteristic functions~\cite{koutrouvelis1981iterative,kogon1998characteristic}. Besides, some recent methods have been developed based on other principles e.g. negative-order moments ~\cite{ma1995blind}, extreme order statistics~\cite{tsihrintzis1996fast} and point process~\cite{marohn1999estimating}.

\begin{figure}[!htbp]
	\centering
	\includegraphics[width=.4\textwidth]{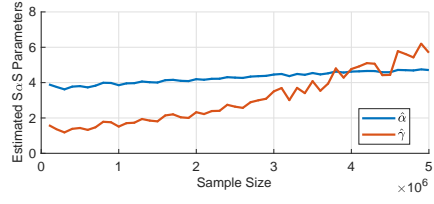}
	\caption{\textit{Tsihrintzis}'s extreme-order-statistic-based estimators in \cite{tsihrintzis1996fast} of \sas~model fail to converge when applied on Bernoulli-Gaussian distributed data. The segmentation parameter of the approach is set to $\left\lfloor\sqrt{N}\right\rfloor$, where $N$ is the sample size.}
	\label{fig:extreme_method_failure}
\end{figure}

As we have derived in Section \ref{sec:stability_bg}, BG processes are not strictly stable. This disqualifies the deployment of some methods listed above on data generated by the BG model, especially the methods that require segmentation of the sample data. For instance, according to our experiment, the extreme-order-statistics-based estimators in \cite{tsihrintzis1996fast} significantly \highlight{depend} on the sample size, and fail to converge, as shown in Fig.~\ref{fig:extreme_method_failure}. 
\begin{figure}[!htbp]
	\centering
	\begin{subfigure}{.49\textwidth}
		\centering
		\includegraphics[width=.9\textwidth]{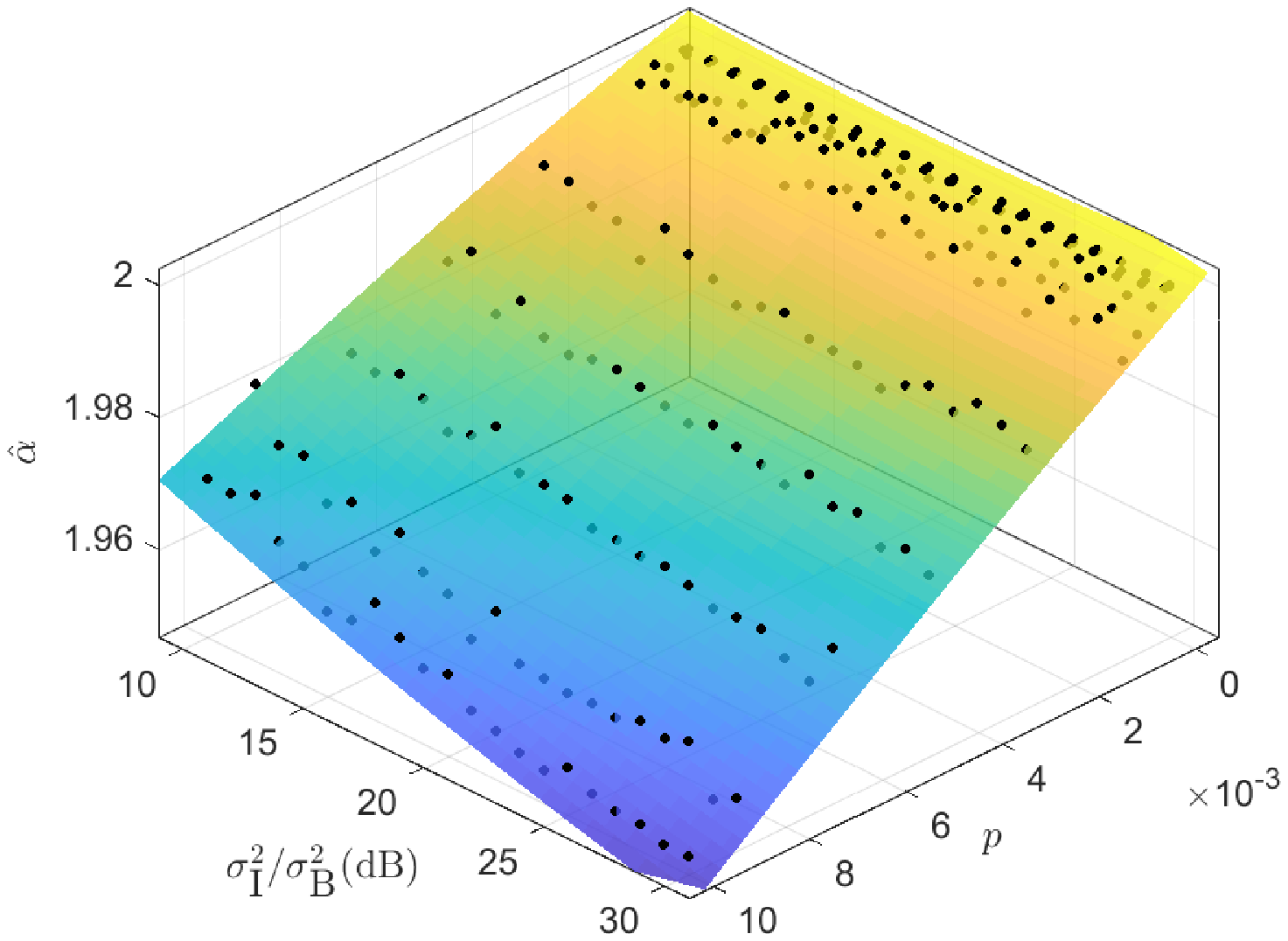}
		\caption{Estimated \sas~parameter $\hat{\alpha}$, with the polynomially fitted surface}
	\end{subfigure}
	\begin{subfigure}{.49\textwidth}
		\centering
		\includegraphics[width=.9\textwidth]{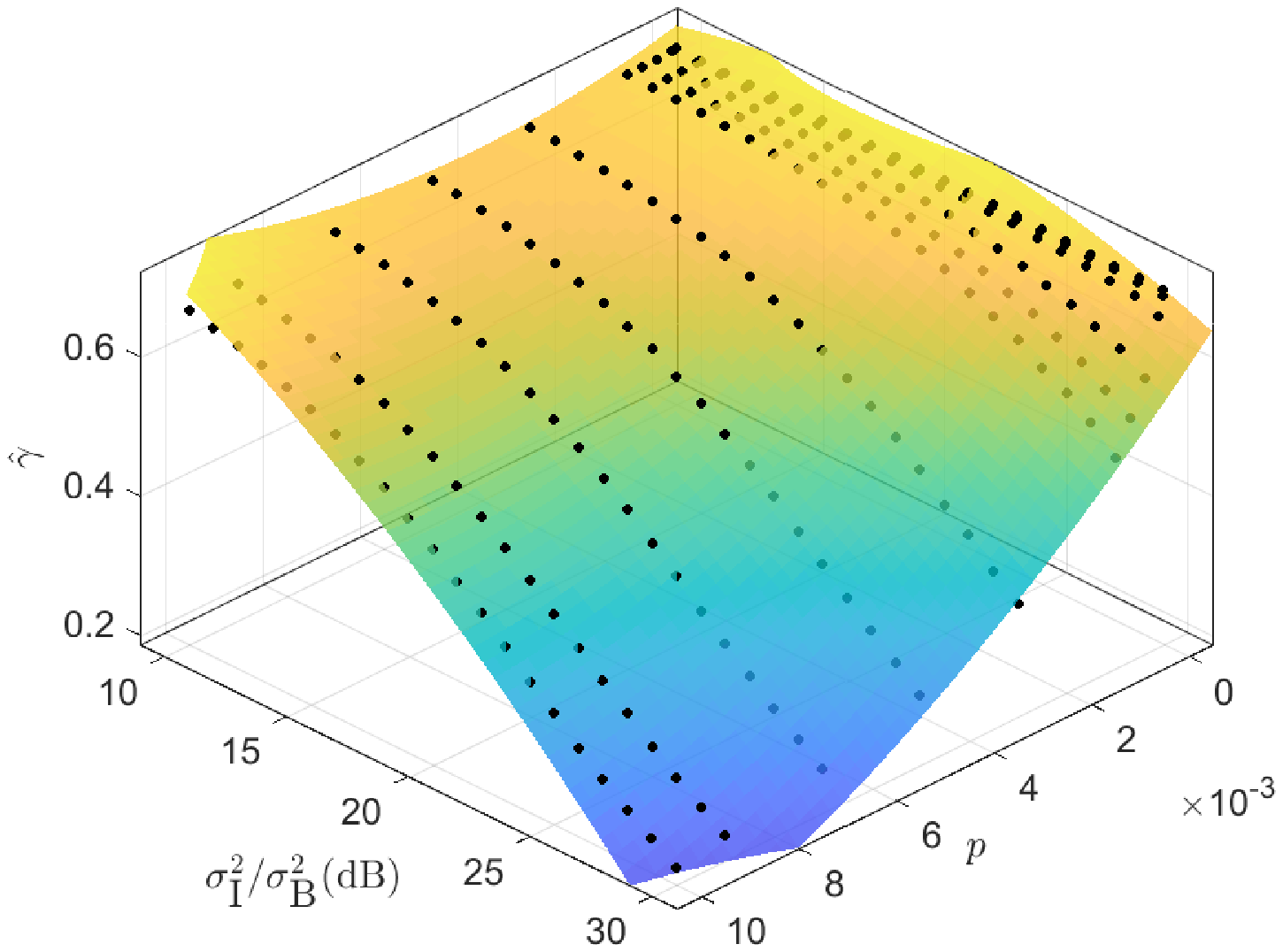}
		\caption{Estimated \sas~parameter $\hat{\gamma}$, with the polynomially fitted surface}
	\end{subfigure}
	\begin{subfigure}{.9\textwidth}
		\centering
		\includegraphics[width=.49\textwidth]{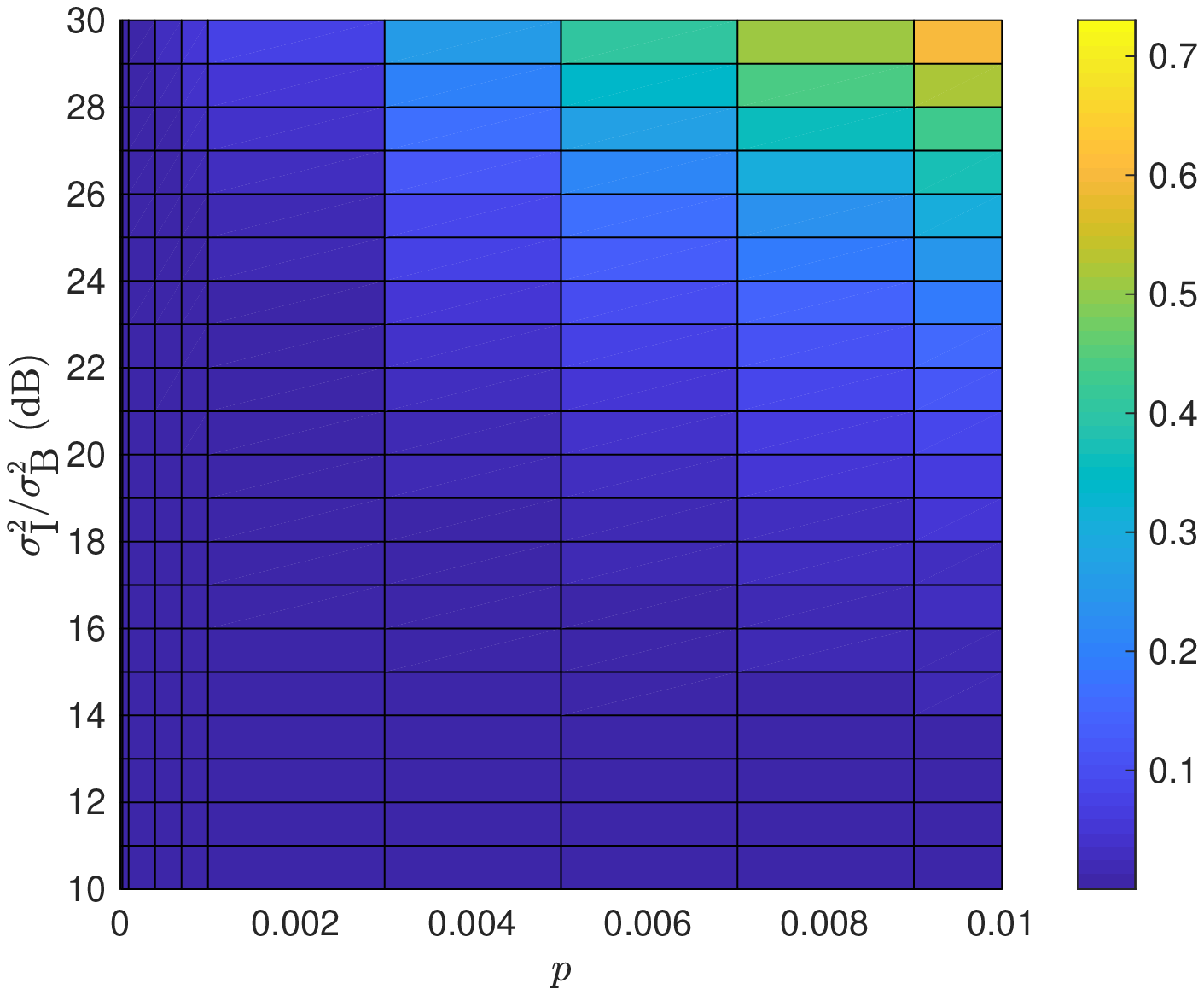}
		\caption{The Kullback-Leibler divergence $D_\textrm{KL}(\phi_\textrm{BG}|\phi_\textrm{\sas})$}
	\end{subfigure}
	\caption{Applying the \sas~model on  Bernoulli-Gaussian noises with different specifications. Under each specification, $5\times10^6$ samples of noise are generated.}
	\label{fig:bg_as_sas}
\end{figure}
Nevertheless, our experiments have proved that at least the regressive methods of {Koutrouvelis} ~\cite{koutrouvelis1980regression,koutrouvelis1981iterative} and {McCulloch} ~\cite{mcculloch1986simple} can be applied, which return similar results. We provide sample outputs obtained by {McCulloch}'s estimators in Section \ref{subsec:parameter_estimation}.

\section{Results and Discussion}\label{sec:results}
\subsection{Fitness Test of Models on Field Measurements}\label{subsec:results_plc_test}
\begin{figure}[!htbp]
	\centering
	\begin{subfigure}{.49\textwidth}
		\centering
		\includegraphics[width=\textwidth]{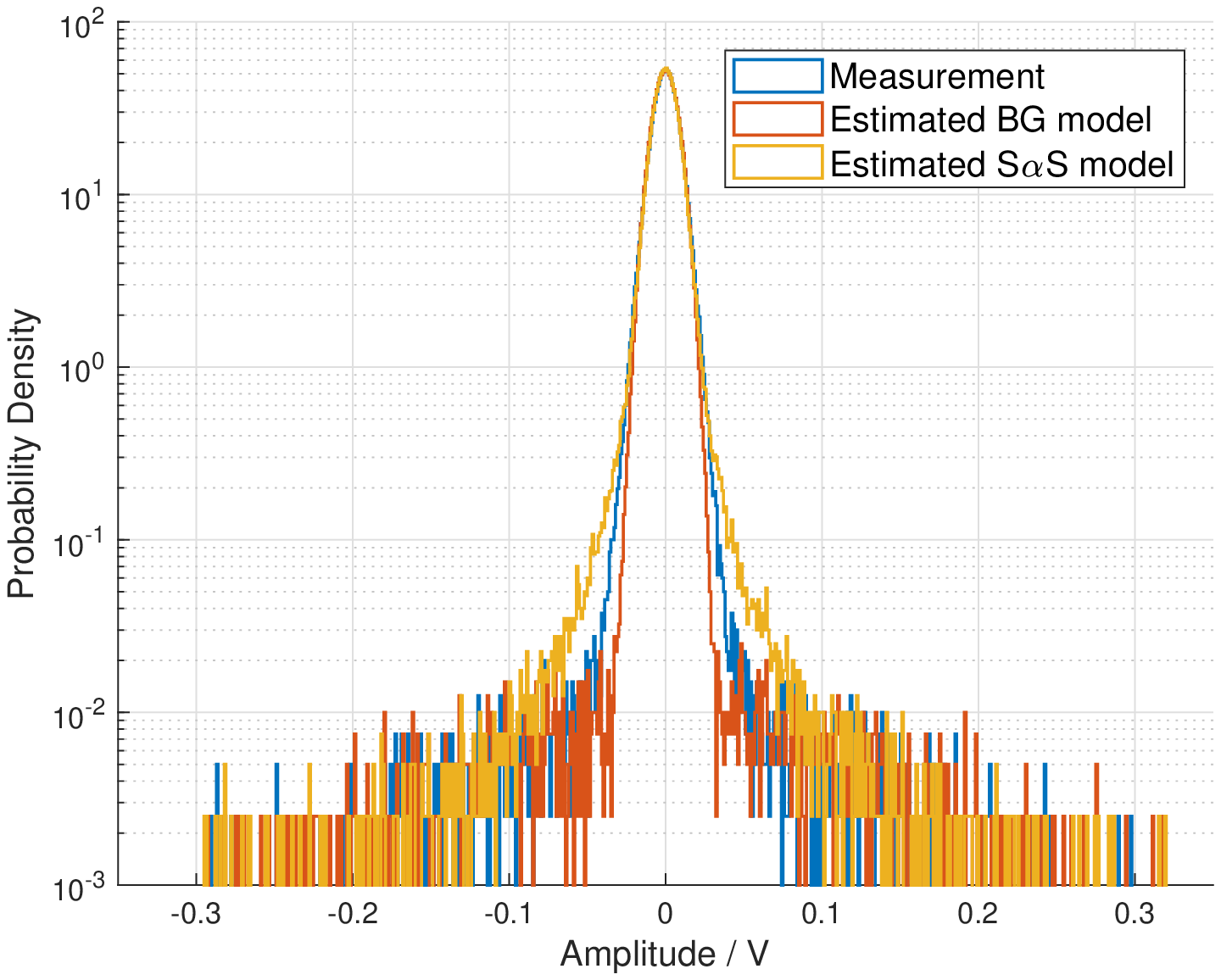}
		\caption{Noise measured on September 9\textsuperscript{th} at 18:17}
	\end{subfigure}
	\begin{subfigure}{.49\textwidth}
		\centering
		\includegraphics[width=\textwidth]{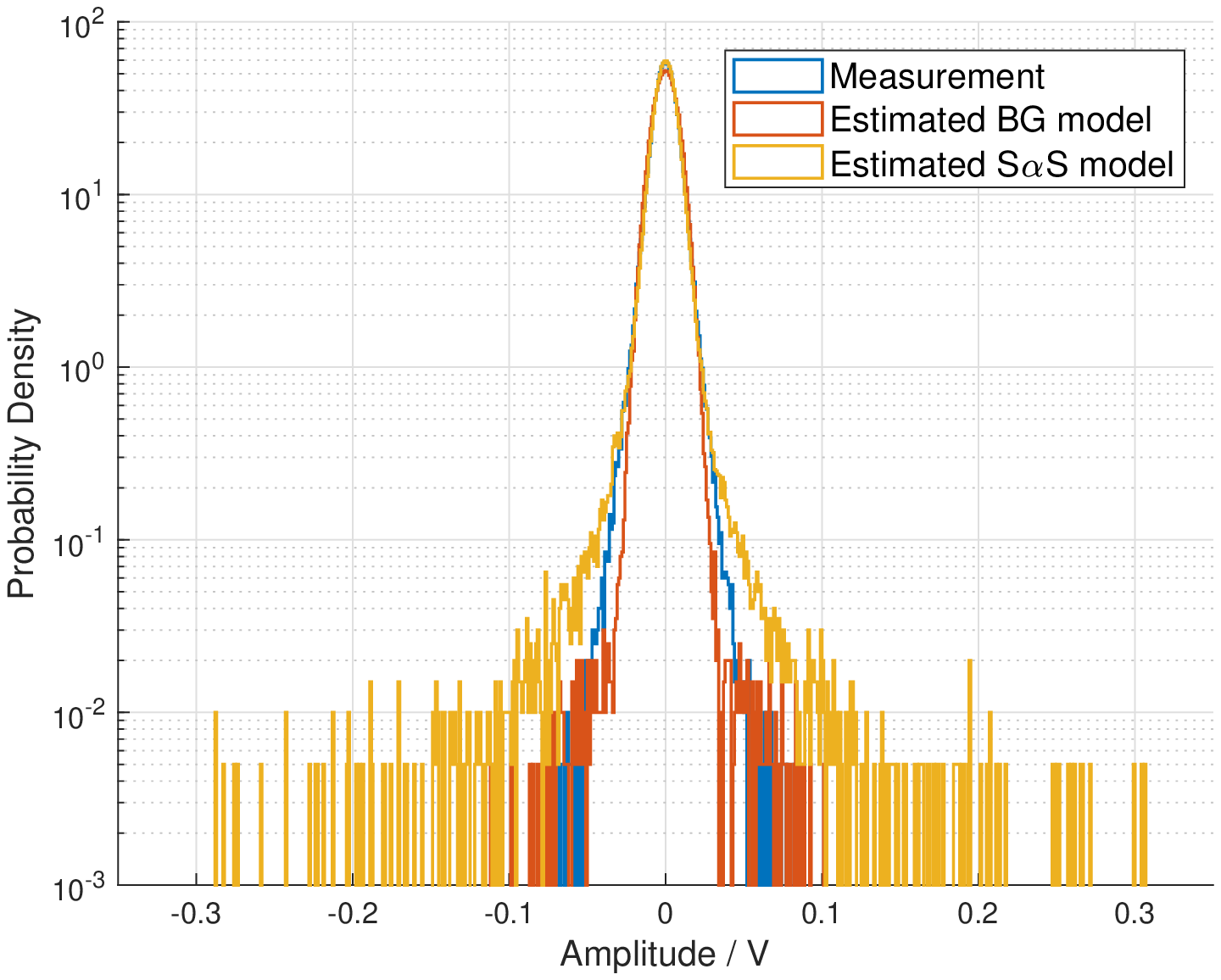}
		\caption{Noise measured on September 14\textsuperscript{th} at 01:08}
	\end{subfigure}
	\caption{Fitting the noise statistics with BG and \sas~models.}
	\label{fig:plc_test}
\end{figure}

The fitting results of BG and \sas~models on field measurements mentioned in Section~\ref{sec:experiment} are illustrated in Fig.~\ref{fig:plc_test}, with the RMSEs listed in Tab.~\ref{tab:plc_fitting_errors}.  It can be summarized that:
\begin{enumerate}
	\item Both BG and \sas~models are able to effectively describe the amplitude distribution of PLC noises, although not perfectly.
	\item When applied on the same PLC noise, the \sas~model gives a wider main lobe in PDF than the measurement, while the BG model has a narrower main lobe than the measurement.
	\item For both models, the fitting performance varies with the noise scenario.
\end{enumerate}

\begin{table}[!htbp]
	\centering
	\caption{Weighted RMSEs of the BG and \sas~models on PLC noise measurements.}
	\label{tab:plc_fitting_errors}
	\begin{tabular}{c|c|c}
	\diaghead{\theadfont \textbf{Model~Measurement}}{\textbf{Fitting Model}}{\textbf{Meas. Time}}&\thead{Sept. 9, 18:17}&\thead{Sept. 14, 01:08}\\\hline
	{BG}&0.0044&0.0019\\\hline
	{\sas}&0.0039&0.0134
	\end{tabular}
\end{table}

\subsection{Stability Test of BG Distribution}\label{subsec:results_bg_stability}
To verify \eqref{equ:quasi_stable_1}--\eqref{equ:quasi_stable_3}, we conducted numerical simulations: \highlight{three} i.i.d. BG random variables $X,Y,Z$ were generated, and another variable $V$ was obtained by $V=(X+Y)\times\sqrt{\frac{\textrm{Var}(Z)}{\textrm{Var}(X+Y)}}$. Then we compared the PDFs of $V$ and $Z$ under different BG model specifications. 
\begin{figure*}[!htpb]
	\centering
	\begin{subfigure}{.49\textwidth}
		\centering
		\includegraphics[width=.85\textwidth]{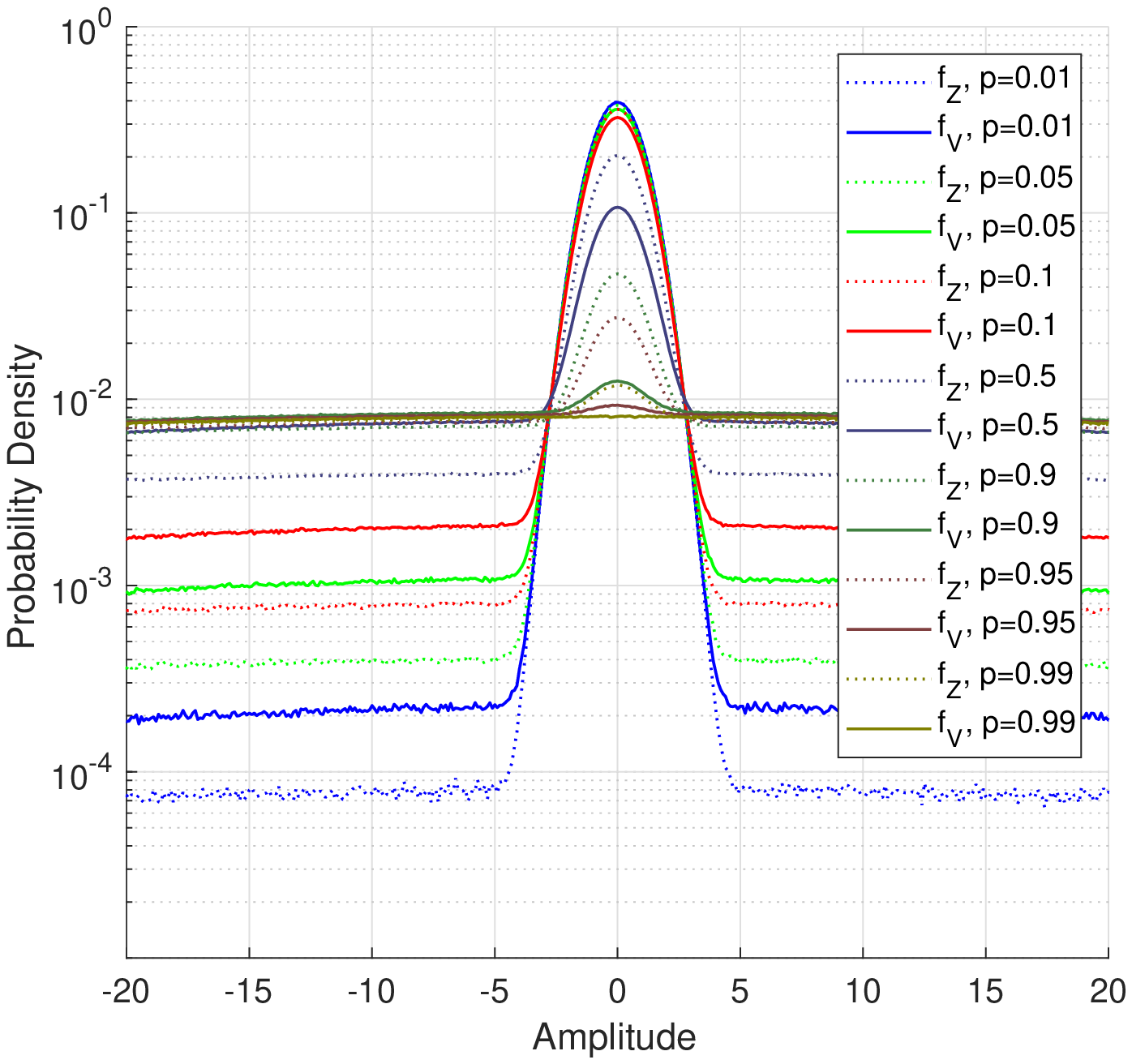}
		\caption{BG distribution with different impulse ratios. $\sigma_\textrm{B}=1$, $\sigma_\textrm{I}=50$.}
		\label{fig:fitness_to_sparsity}
	\end{subfigure}
	\begin{subfigure}{.49\textwidth}
		\centering
		\includegraphics[width=.85\textwidth]{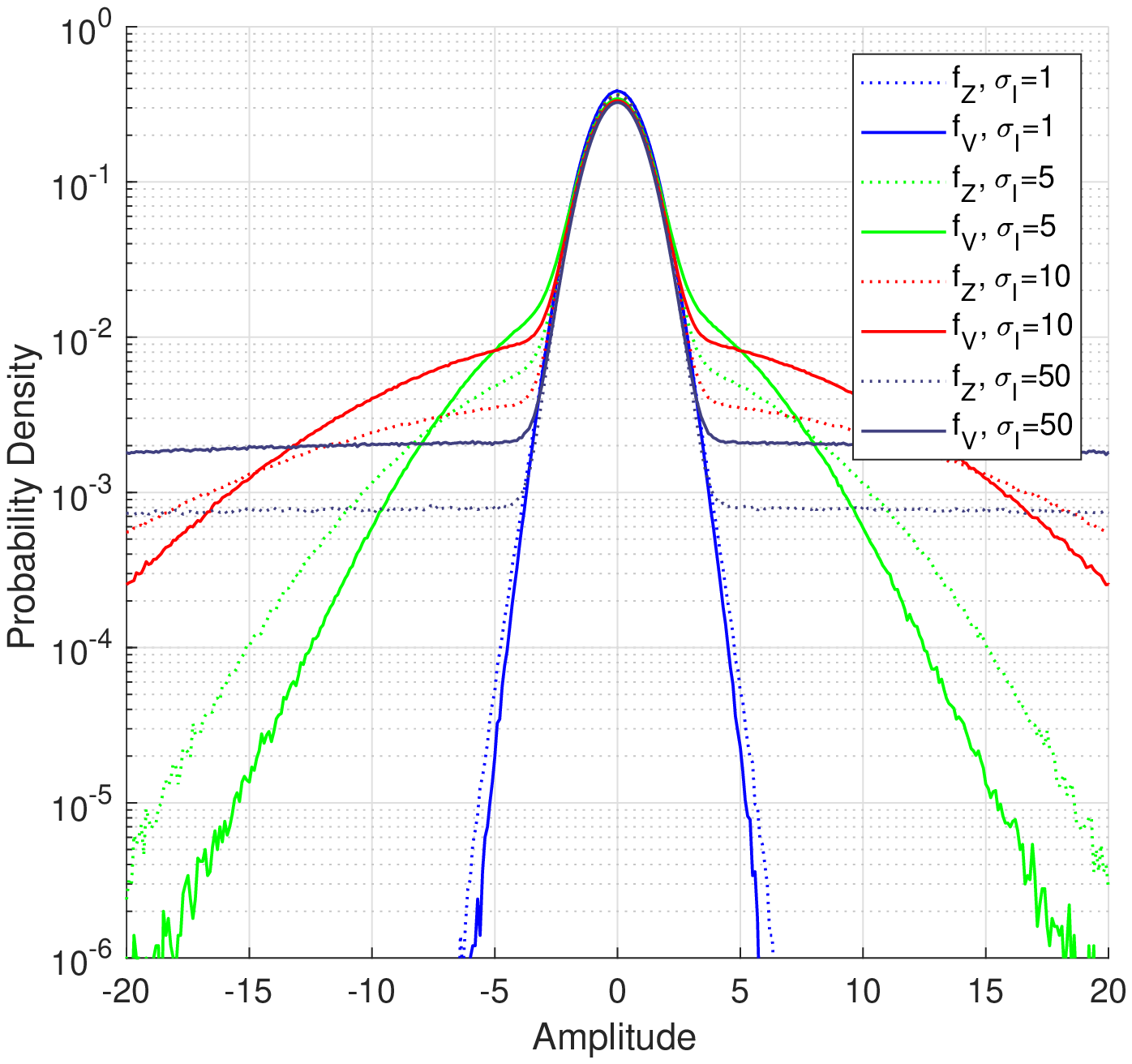}
		\caption{BG distribution with different impulse deviations. $p=0.1$, $\sigma_\textrm{B}=1$.}
		\label{fig:fitness_to_par}
	\end{subfigure}
	\caption{Testing the stability of BG distribution with respect to the model parameters.}
\end{figure*}
First, the deviations were fixed to $\sigma_\textrm{B}=1$ and $\sigma_\textrm{I}=50$, and the test was executed for different values of $p$. The results are shown in Fig.~\ref{fig:fitness_to_sparsity}. It can be observed that the PDFs of $V$ and $Z$ match each other well when the value of $p$ is \highlight{sufficiently} high or low, but deviate from each other as $p$ approaches to $0.5$, which matches our theory. Subsequently, we set $p=0.1$, $\sigma_\textrm{B}^2=1$, and repeated the test for different values of $\sigma_\textrm{I}$. The results in Fig.~\ref{fig:fitness_to_par} show that $V$ has its  distribution more similar to $Z$ under lower impulse power, as we have expected.

In the context of PLC, according to ~\cite{zimmermann2000analysis}, over the broad band up to \SI{20}{\mega\hertz}, the impulse power is usually by \SIrange{10}{30}{\dB}, i.e., 10 to 1000 times higher than that of the background noise. Meanwhile, the impulse probability remains below $0.35\%$ even under the heaviest disturbance, and falls down to $0.00135\%$ under the weak disturbance. Comparing these to the result in Fig.~\ref{fig:fitness_to_sparsity}, it is reasonable to consider the BG model as quasi-stable when applied on power line noises.

\subsection{\sas~Parameter Estimation of BG Processes}\label{subsec:parameter_estimation}
Setting $\sigma_\textrm{B}=1$ with different values of $0.01\%\le p\le1\%$ and $\SI{10}{\dB}\le\frac{\sigma_\textrm{I}^2}{\sigma_\textrm{B}^2}\le\SI{30}{\dB}$, we applied {McCulloch}'s estimators on randomly generated BG processes, and got the results in Fig.~\ref{fig:bg_as_sas}.  We can observe from the Kullback-Leibler divergence that the conversion can successfully provide a precise approximation when either the impulse-to-background power ratio $\frac{\sigma_\textrm{I}^2}{\sigma_\textrm{B}^2}$ or the impulse ratio $p$ is limited. In the cases with extremely intensive impulses where both $\frac{\sigma_\textrm{I}^2}{\sigma_\textrm{B}^2}$ and $p$ are high, the fitness of model conversion sinks dramatically. We also fitted (2,2)-polynomial surfaces for $\hat{\alpha}$ and $\hat{\gamma}$ to support fast and approximate conversions from power-normalized BG model to \sas~model, the results are listed in Tab.~\ref{tab:bg_as_sas_fitting}.


\begin{table}[!htbp]
	\centering
	\caption{(2,2)-polynomial surface fitting results of $\hat{\alpha}$ and $\hat{\gamma}$ as function of $p$ and $\frac{\sigma_\textrm{I}}{\sigma_\textrm{B}}$.}
	\label{tab:bg_as_sas_fitting}
	\begin{tabular}{c|c|c}
		&$\hat{\alpha}$&$\hat{\gamma}$\\\hline
		$c_{00}$ &2.005&0.5779\\\hline
		$c_{10}$ &-1.457&6.256\\\hline
		$c_{01}$ &$-5.575\times10^{-4}$&0.01707\\\hline
		$c_{20}$ &-40.36&2123\\\hline
		$c_{11}$ &-0.1128&-2.43\\\hline
		$c_{02}$ &$1.426\times10^{-5}$&$-5.249\times10^{-4}$\\\hline
		RMSE &$1.715\times 10^{-3}$&$2.433\times10^{-2}$\\\hline
		\multicolumn{3}{l}{Fitting model: $f(x,y)=c_{00}+c_{10}x+c_{01}y+c_{20}x^2$ }\\
		\multicolumn{3}{l}{$+c_{11}xy+c_{02}y^2$, where $x=p$, $y=20\log_{10}(\sigma_\textrm{I}/\sigma_\textrm{B})$}
	\end{tabular}
\end{table}

\section{Conclusion}\label{sec:conclusion}
In this paper, we have studied the compatibility between two widely-used models for impulsive power line noises: the BG model and the \sas~model. With field measurement test, we have proved that they give different but similarly acceptable results when used to model power line noise. Then we have \highlight{proved} that the BG distribution is neither strictly stable nor strictly fat-tailed, so that no analytical unification between BG and \sas~models is feasible. Nevertheless, when the impulses are sparse and not extremely strong in power, which is the common case of power line noises, BG processes can be approximately considered as quasi-stable, so that an approximate and empirical model conversion is possible. Based on this result, we have proposed a fast  (2,2)-polynomial conversion from BG model to \sas~model. This fast conversion can be applied to merge reference power line noise scenarios based on different models, and hence to simplify the performance evaluation of PLC systems.

\appendix
\section{The PDF of Sum of Two I.I.D. BG Noises}
The detailed derivation of \eqref{equ:bg_sum_pdf} follows below:

\begin{align}
\begin{split}
&f_W(w)=\int\limits_{-\infty}^{+\infty}f_\textrm{BG}(w-x)f_\textrm{BG}(x)\mathrm{d}x\\
=&\int\limits_{-\infty}^{+\infty}\left[\frac{1-p}{\sqrt{2\pi\sigma_\textrm{B}^2}}\mathrm{e}^{-\frac{(w-x)^2}{2\sigma_\textrm{B}^2}}+\frac{p}{\sqrt{2\pi(\sigma_\textrm{B}^2+\sigma_\textrm{I}^2)}}\mathrm{e}^{-\frac{(w-x)^2}{2(\sigma_\textrm{B}^2+\sigma_\textrm{I}^2)}}\right]\\
&\times\left[\frac{1-p}{\sqrt{2\pi\sigma_\textrm{B}^2}}\mathrm{e}^{-\frac{x^2}{2\sigma_\textrm{B}^2}}+\frac{p}{\sqrt{2\pi(\sigma_\textrm{B}^2+\sigma_\textrm{I}^2)}}\mathrm{e}^{-\frac{x^2}{2(\sigma_\textrm{B}^2+\sigma_\textrm{I}^2)}}\right]\mathrm{d}x\\
=&\int\limits_{-\infty}^{+\infty}\left[\frac{(1-p)^2}{2\pi\sigma_\textrm{B}^2}\mathrm{e}^{-\frac{(w-x)^2+x^2}{2\sigma_\textrm{B}^2}}+\frac{p^2}{2\pi(\sigma_\textrm{B}^2+\sigma_\textrm{I}^2)}\mathrm{e}^{-\frac{(w-x)^2+x^2}{2(\sigma_\textrm{B}^2+\sigma_\textrm{I}^2)}}\right. \\
&\left.+\frac{p(1-p)}{2\pi\sigma_\textrm{B}\sqrt{\sigma_\textrm{B}^2+\sigma_\textrm{I}^2}}\left(\mathrm{e}^{-\frac{(w-x)^2}{2(\sigma_\textrm{B}^2+\sigma_\textrm{I}^2)}-\frac{x^2}{2\sigma_\textrm{B}^2}}+\mathrm{e}^{-\frac{x^2}{2(\sigma_\textrm{B}^2+\sigma_\textrm{I}^2)}-\frac{(w-x)^2}{2\sigma_\textrm{B}^2}}\right)\right]\mathrm{d}x\\
=&\int\limits_{-\infty}^{+\infty}\frac{(1-p)^2}{2\pi\sigma_\textrm{B}^2}\mathrm{e}^{-\frac{2x^2-2wx+w^2}{2\sigma_\textrm{B}^2}}\mathrm{d}x+\int\limits_{-\infty}^{+\infty}\frac{p^2}{2\pi(\sigma_\textrm{B}^2+\sigma_\textrm{I}^2)}\mathrm{e}^{-\frac{2x^2-2wx+w^2}{2(\sigma_\textrm{B}^2+\sigma_\textrm{I}^2)}}\mathrm{d}x\\
+&\int\limits_{-\infty}^{+\infty}\frac{p(1-p)}{2\pi\sigma_\textrm{B}\sqrt{\sigma_\textrm{B}^2+\sigma_\textrm{I}^2}}\left(\mathrm{e}^{-\frac{(2\sigma_\textrm{B}^2+\sigma_\textrm{I}^2)x^2-2\sigma_\textrm{B}^2wx+\sigma_\textrm{B}^2w^2}{2\sigma_\textrm{B}^2(\sigma_\textrm{B}^2+\sigma_\textrm{I}^2)}}\right.\\
&\left.+\mathrm{e}^{-\frac{(2\sigma_\textrm{B}^2+\sigma_\textrm{I}^2)x^2-2(\sigma_\textrm{B}^2+\sigma_\textrm{I}^2)wx+(\sigma_\textrm{B}^2+\sigma_\textrm{I}^2)w^2}{2\sigma_\textrm{B}^2(\sigma_\textrm{B}^2+\sigma_\textrm{I}^2)}}\right)\mathrm{d}x.
\end{split}
\end{align}

Let $\sigma_1=\sqrt{\sigma_\textrm{B}^2+\sigma_\textrm{I}^2}$, $\sigma_2=\sqrt{2\sigma_\textrm{B}^2+\sigma_\textrm{I}^2}=\sqrt{\sigma_\textrm{B}^2+\sigma_1^2}$:

\begin{align}
\begin{split}
f_W(w)=&\frac{(1-p)^2}{2\pi\sigma_\textrm{B}^2}\int\limits_{-\infty}^{+\infty}\mathrm{e}^{-\frac{2x^2-2wx+w^2}{2\sigma_\textrm{B}^2}}\mathrm{d}x+\frac{p^2}{2\pi\sigma_1^2}\int\limits_{-\infty}^{+\infty}\mathrm{e}^{-\frac{2x^2-2wx+w^2}{2\sigma_1^2}}\mathrm{d}x\\
&+\frac{p(1-p)}{2\pi\sigma_\textrm{B}\sigma_1}\int\limits_{-\infty}^{+\infty}\left(\mathrm{e}^{-\frac{\sigma_2^2x^2-2\sigma_\textrm{B}^2wx+\sigma_\textrm{B}^2w^2}{2\sigma_\textrm{B}^2\sigma_1^2}}+\mathrm{e}^{-\frac{\sigma_2^2x^2-2\sigma_1^2wx+\sigma_1^2w^2}{2\sigma_\textrm{B}^2\sigma_1^2}}\right)\mathrm{d}x\\
=&\frac{1-2p+p^2}{\sqrt{4\pi\sigma_\textrm{B}^2}}\mathrm{e}^{-\frac{w^2}{4\sigma_\textrm{B}^2}}\int\limits_{-\infty}^{+\infty}\frac{1}{\sqrt{2\pi\sigma_\textrm{B}^2}}\mathrm{e}^{-\frac{\left(\sqrt{2}x-\frac{\sqrt{2}}{2}w\right)^2}{2\sigma_\textrm{B}^2}}\mathrm{d}\left(\sqrt{2}x-\frac{\sqrt{2}}{2}w\right)\\
&+\frac{p^2}{\sqrt{4\pi\sigma_1^2}}\mathrm{e}^{-\frac{w^2}{4\sigma_1^2}}\int\limits_{-\infty}^{+\infty}\frac{1}{\sqrt{2\pi\sigma_1^2}}\mathrm{e}^{-\frac{\left(\sqrt{2}x-\frac{\sqrt{2}}{2}w\right)^2}{2\sigma_1^2}}\mathrm{d}\left(\sqrt{2}x-\frac{\sqrt{2}}{2}w\right)\\
+&\frac{p-p^2}{\sqrt{2\pi\sigma_2^2}}\mathrm{e}^{-\frac{(\sigma_2^2-1)w^2}{2\sigma_1^2\sigma_2^2}}\int\limits_{-\infty}^{+\infty}\frac{1}{\sqrt{2\pi\sigma_\textrm{B}^2\sigma_1^2}}\mathrm{e}^{-\frac{\left(\sigma_2x-\frac{\sigma_\textrm{B}}{\sigma_2}w\right)^2}{2\sigma_\textrm{B}^2\sigma_1^2}}\textrm{d}\left(\sigma_2x-\frac{\sigma_\textrm{B}}{\sigma_2}w\right)\\
&+\frac{p-p^2}{\sqrt{2\pi\sigma_2^2}}\mathrm{e}^{-\frac{(\sigma_2^2-1)w^2}{2\sigma_\textrm{B}^2\sigma_2^2}}\int\limits_{-\infty}^{+\infty}\frac{1}{\sqrt{2\pi\sigma_\textrm{B}^2\sigma_1^2}}\mathrm{e}^{-\frac{\left(\sigma_2x-\frac{\sigma_1}{\sigma_2}w\right)^2}{2\sigma_\textrm{B}^2\sigma_1^2}}\textrm{d}\left(\sigma_2x-\frac{\sigma_1}{\sigma_2}w\right).
\end{split}
\label{equ:derivation_middle_stage}
\end{align}
\normalsize
Note that all the integration terms in the last step of \eqref{equ:derivation_middle_stage} follow the form of normal distribution. Hence, all the integrations over $(-\infty,\infty)$ have the value of 1, and we get:

\begin{align}
\begin{split}
f_\textrm{W}(w)=&\frac{1-2p+p^2}{\sqrt{4\pi\sigma_\textrm{B}^2}}\mathrm{e}^{-\frac{w^2}{4\sigma_\textrm{B}^2}}+\frac{p^2}{\sqrt{4\pi\sigma_1^2}}\mathrm{e}^{-\frac{w^2}{4\sigma_1^2}}\\
+&\frac{p-p^2}{\sqrt{2\pi\sigma_2^2}}\left(\mathrm{e}^{-\frac{(\sigma_2^2-1)w^2}{2\sigma_1^2\sigma_2^2}}+\mathrm{e}^{-\frac{(\sigma_2^2-1)w^2}{2\sigma_\textrm{B}^2\sigma_2^2}}\right).
\end{split}
\end{align}

\bibliographystyle{elsarticle-num}
\bibliography{references}
\end{document}